%%%%%%%%%%%%%%%%%%%%%%%%%%%%%%%%%%%%%%%%%%%%%%%%%%%%%%%%%%%%%%%%%%%%%%%%
%% Character Table
%%   Upper-case     A B C D E F G H I J K L M N O P Q R S T U V W X Y Z
%%   Lower-case     a b c d e f g h i j k l m n o p q r s t u v w x y z
%%   Digits         0 1 2 3 4 5 6 7 8 9
%%   Exclamation    !     Double quote   "     Hash (number)  #
%%   Dollar         $     Percent        %     Ampersand      &
%%   Acute accent   '     Left paren     (     Right paren    )
%%   Asterisk       *     Plus           +     Comma          ,
%%   Minus          -     Point          .     Solidus        /
%%   Colon          :     Semicolon      ;     Less than      <
%%   Equals         =     Greater than   >     Question mark  ?
%%   Commercial at  @     Left bracket   [     Backslash      \ 
%%   Right bracket  ]     Circumflex     ^     Underscore     _
%%   Grave accent   `     Left brace     {     Vertical bar   |
%%   Right brace    }     Tilde          ~
%%%%%%%%%%%%%%%%%%%%%%%%%%%%%%%%%%%%%%%%%%%%%%%%%%%%%%%%%%%%%%%%%%%%%%%%
\magnification=\magstep 1
\tolerance=1600
\font\ner=cmbx10
\font\tit=cmr10 
\baselineskip= 0.85 true cm
\vsize= 20 true cm
\topinsert 
\vskip 50pt
\endinsert

{\ner
\centerline{Low Dimensional Ordering on a Lattice Model}}

\vskip 15pt

{\tit
\hfill Fabio Siringo\hfill}

\vskip 5pt

{\tit
\hfill Dipartimento di Fisica dell'Universit\`a di Catania \hfill}

{\tit
\hfill Corso Italia 57, I 95129 Catania - Italy\hfill}

\vskip 30pt

\centerline{\ner
Abstract }

\vskip 15pt

{\tit\narrower{
A simple $d$-dimensional lattice model is proposed, incorporating
some degree of frustration and thus capable of describing some
aspects of molecular orientation in covalently bound molecular solids.
For $d=2$ the model is shown to be equivalent to the standard 
two-dimensional Ising model, while for $d>2$ it describes a peculiar
transition from an isotropic high temperature phase to a low-dimensional
anisotropic low temperature state. A general mean field analysis is
presented and compared to some exact limit properties.

\smallskip}}

\vskip 15pt

{\tit
}

{\tit
}

\vskip 15pt

{\tit
E-mail address: siringo@ct.infn.it\par
Fax: ++39-95-383023}

\vfill\eject

{\tit
Most molecular liquids retain their molecular structure even in
the solid phase, where some long range order usually shows up as
a consequence of inter-molecular interaction. However in the solid
the orientational order of the molecules may change according to
the thermodynamic conditions giving rise to quite rich phase diagrams
as recently observed for hydrogen under high pressure [1]. Orientational 
ordering is responsible for several phase transitions occurring even
in the liquid phase (liquid crystals) and is thought to be related
to the metal-insulator transition of liquid iodine [2]. 

Molecular ordering may be classified into two different classes
according to the nature of the interaction. The usual interaction
has a quadrupole nature, is weak and gives rise to the observed
three-dimensional ordering of most molecular Van der Waals solids.
Conversely in some molecular solids the interaction has a covalent
main component and is characterized by some level of frustration
since the coordination number for the covalent bond is quite low.
Each molecule must choose a few partners and cannot accept any
further proposal. The lower is the allowed coordination number,
the higher the frustration which gives rise to a low-dimensional
structure as observed in polymers (one-dimensional) or in solid
iodine [3,4] (two-dimensional). Actually some one-dimensional zigzag
chains have also been reported [5] in iodine where the covalent nature
of the interaction is out of doubt [6,7]. Moreover we expect that a
covalent interaction should show up for all the molecular solids
under high pressure, as the inter-molecular distance approaches the
intra-molecular length, provided that some important structural
transition does not occur first (as dissociation).

While the thermodynamic behaviour of the quadrupole driven ordering
can be analyzed in terms of $O(3)$ symmetric vectorial models, in
this letter we introduce a simple frustrated lattice model which
is capable of describing some aspects of a covalently bound molecular
solid. More precisely the model describes the transition from an
high-temperature (or weakly interacting) fully isotropic disordered
system, to a low-temperature (or strongly interacting) anisotropic
 low dimensional broken-symmetry phase. As a consequence of frustration
 the breaking of symmetry is accompanied by a sort of decomposition
 of the system in low-dimensional almost independent parts, as observed
 in solid iodine. Such remarkable behaviour requires a space dimension
 $d>2$, while for $d=2$ the model is shown to be equivalent 
 to the exactly solvable two-dimensional Ising model [8]. Then, 
 apart from its original motivation, the model seems to be 
 of interest by itself as an unconventional generalization
 of the standard Ising model to higher dimensions.

 Let us consider a $d$-dimensional hypercubic lattice, with a
 randomly oriented linear molecule at each site. The molecules are
 supposed to be symmetric with respect to their centre of mass
 which is fixed at the lattice site. The description is 
 simplified by allowing only a discrete number of space orientations
 for each molecule: we assume that each of them must point towards
 one of its $2d$ first neighbour sites. This choice can be justified
 by the existence of covalent interactions along preferred axes.
 In this {\it quantized} version each molecule has $d$ different
 states corresponding to molecular orientation along the hypercube
 axes (molecules are supposed to be symmetric). Finally, each couple
 of first neighbour molecules, when pointing the one towards the other,
 are assumed to gain a bonding energy for their directional covalent
 bond (they touch each other). As shown in figure 1 for $d=2$, bonding
 in a direction excludes any possible bond along the other $(d-1)$
 directions. The coordination number is $2$ for any value of $d$, and
 the frustration increases with increasing $d$.

 According to such description we introduce a variable versor 
 $\hat w_{\bf r}$ for each of the $N$ sites ${\bf r}$ of the lattice,
 with $\hat w_{\bf r} \in\{\hat x_1, \hat x_2,\cdots \hat x_d\}$ 
 pointing towards one of the $d$ hypercube axes $\hat x_\alpha$.
 The partition function follows
 $$Z=\sum_{\{\hat w\}} e^S=
 \sum_{\{\hat w\}} e^{4\beta \sum_{{\bf r},\alpha} 
 \delta_{\hat w_{\bf r} \hat x_\alpha} \cdot 
 \delta_{\hat w_{{\bf r}+\hat x_\alpha} \hat x_\alpha}} \eqno(1)$$ 
 where $\{\hat w\}$ indicates a sum over all configurations,
 $\alpha$ runs from $1$ to $d$, and the lattice spacing is set to
 unity. The inverse temperature $\beta$, in units of binding energy,
 can be negative for a {\it repulsive} model, but is assumed positive
 in the present molecular context.

 We first show the equivalence for $d=2$ between molecular and Ising
 model. Let us add for each couple of sites a generic field 
 ${\bf h}(\alpha)$  according to the space direction $\alpha$ of the
 lattice link which joins the sites. The partition function reads
 $$Z_h=\sum_{\{\hat w\}} e^{S_h}=
 \sum_{\{\hat w\}} e^{\displaystyle 
 \>\>4\beta \sum_{{\bf r},\alpha}\left( 
 \delta_{\hat w_{\bf r} \hat x_\alpha} \cdot 
 \delta_{\hat w_{{\bf r}+\hat x_\alpha} \hat x_\alpha}
 + {\bf h}(\alpha)\hat w_{\bf r}
 + {\bf h}(\alpha)\hat w_{{\bf r}+\hat x_\alpha}\right)} \eqno(2)$$ 
Introducing a matrix notation
$$Z_h=\sum_{\{\hat w\}} e^{S_h}=\sum_{\{\hat w\}} 
e^{\sum_{{\bf r},\alpha} L({\bf r},\alpha)}\eqno(3)$$
the Lagrangian density $L$ follows as
$$L({\bf r},\alpha)= \hat w_{\bf r}^\dagger M_\alpha(\beta, {\bf h})
\hat w_{{\bf r}+\hat x_\alpha} \eqno(4)$$
Here the canonical $d$-dimensional column vector representation  
of $R^d$ is employed with $\hat x_1 \equiv (1,0,0\dots)$,
$\hat x_2 \equiv (0,1,0\dots)$, etc. The $d\times d$ matrix $M_\alpha$
does not depend on the configurations of the system, and entirely
characterizes it. Now provided that $\sum_\alpha {\bf h}(\alpha)=0$,
$S_h$ does not depend on ${\bf h}$ and $S_h\equiv S$. For $d=2$
such condition is satisfied by the field ${\bf h} (1)=h(\hat x_1
-\hat x_2)$, ${\bf h} (2)=-{\bf h}(1)$. The matrix $M_\alpha$ follows
$$M_1=\left(\matrix{4\beta(1+2h) & 0 \cr 
             0            & -8\beta h\cr}\right)\>;\>
M_2=\left(\matrix{-8\beta h & 0 \cr 
                   0        & 4\beta (1+2h) \cr}\right)\eqno(5)$$
Then for $h=-1/4$, $M_1\equiv M_2$, and $L$ reads
$$L({\bf r}, \alpha)=\beta+\hat w_{\bf r}^\dagger
\left(\matrix{\beta & -\beta \cr -\beta & \beta\cr}\right)
\hat w_{{\bf r}+\hat x_\alpha}.\eqno(6)$$
Identifying the two-dimensional versors $\hat w$ with spin variables,
apart from an inessential factor, $Z$ reduces to the partition
function of a two-dimensional Ising model
$$Z=e^{\beta N d}\cdot Z_{Ising}\eqno(7)$$
and is exactly solvable.
For $\beta \to +\infty$ a ground state is approached with all the
molecules aligned along the same direction, and formation of
one-dimensional polymeric chains (figure 2a); for $\beta \to
-\infty$ the repulsive model approaches a zero-energy (no bonds)
ground state analogous to the antiferromagnetic configuration of
the Ising model (figure 2b).

For $d\ge 3$ the analogy with the Ising model breaks down, and this
is evident from a simple analysis of the ground state configuration.
Due to frustration the model has an infinitely degenerate ground state
in the thermodynamic limit $N\to \infty$. For instance in the case
$d=3$, the minimum energy is obtained by aligning all the molecules
along a common direction, as for $d=2$. However the ground state
configuration is not unique: the number of molecular bonds does not
change if we rotate together all the molecules belonging to an entire
layer which is parallel to the original direction of alignment.
As a consequence of frustration the total degeneration is
$3(2^{(N^{1/3})})$, and the system is expected to behave like a glass
for the large energy threshold which separate each minimum from the
other. The phase diagram is expected to be quite rich, with at least
a transition point between the high temperature disordered phase
and an ordered broken-symmetry low temperature phase. 

Some analytical result can be obtained in Mean-Field (MF) approximation:
neglecting second order fluctuation terms
$$\delta_{\hat w_{\bf r} \hat x_\alpha} \cdot 
\delta_{\hat w_{{\bf r}+\hat x_\alpha} \hat x_\alpha}\approx
\delta_{\hat w_{\bf r} \hat x_\alpha} \cdot  \Delta_\alpha+
\delta_{\hat w_{{\bf r}+\hat x_\alpha} \hat x_\alpha}\cdot \Delta_\alpha
-\Delta_\alpha^2\eqno(8)$$
where $\Delta_\alpha=\langle 
\delta_{\hat w_{\bf r} \hat x_\alpha}\rangle$, and $\sum_\alpha
\Delta_\alpha=1$ (with the constraint $0\le\Delta_\alpha\le 1$).
The partition function factorizes as
$$Z_{MF}=e^{-4N\beta\sum_\alpha \Delta_\alpha^2}
\left[\sum_\alpha e^{8\beta\Delta_\alpha}\right]^N\eqno(9)$$
and the free energy follows
$$F_{MF}=-{1\over{N\beta}} \log Z_{MF}=4\sum_\alpha \Delta_\alpha^2
-{1\over\beta}\log\left(\sum_\alpha e^{8\beta\Delta_\alpha}\right).
\eqno(10)$$
The derivative with respect to $\Delta_\mu$ yields, for the stationary
points
$$\Delta_\mu={{e^{8\beta\Delta_\mu}}\over{\sum_\alpha e^{8\beta\Delta
_\alpha}}}\eqno(11)$$
which satisfies the condition $\sum_\alpha\Delta_\alpha=1$.

In the high temperature limit $\beta\to 0$ eq. (11) admits the unique
solution $\Delta_\mu=1/d$ which reflects the complete random orientation
of molecules. In the opposite limit $\beta\to\infty$, apart from such
solution, eq. (11) is satisfied by the broken-symmetry field 
$\Delta_\mu=1$, $\Delta_\alpha=0$ for $\alpha\not=\mu$, which obviously
corresponds to a minimum for $F_{MF}$. Then at a critical point 
$\beta=\beta_c$ the high temperature solution must become unstable
towards a multivalued minimum configuration.
The Hessian matrix is easily evaluated at the stationary points
employing eq. (11)
$$H_{\mu \nu}={1\over 8} {{\partial^2 F_{MF}}\over{\partial\Delta_\mu
\partial\Delta_\nu}}=\delta_{\mu \nu}\left(1-8\beta\Delta_\mu\right)
+8\beta\Delta_\mu\Delta_\nu\eqno(12)$$
For $\beta<\beta_c$, inserting $\Delta_\mu=1/d$, the
eigenvalue problem 
$det\vert H_{\mu\nu}-\lambda\delta_{\mu\nu}\vert=0$
yields
$$\left(1-{{8\beta}\over{d}}-\lambda\right)^{d-1}\cdot\left(1-\lambda
\right)=0.\eqno(13)$$
Thus the Hessian matrix is positive defined if and only if         
$\lambda=(1-8\beta/d)>0$. Beyond the critical point $\beta=\beta_c=
(d/8)$ the solution $\Delta_\mu=1/d$ is not a minimum, and a multivalued
minimum configuration shows up. Such result obviously agrees with the
MF prediction for the Ising model $\beta_{Ising}=1/(2d)$, only for
the special dimension $d=2$. For $d>2$ we observe an increase of
$\beta_c$ with $d$, to be compared to the opposite trend shown by the
Ising model. Such behaviour may be interpreted in terms of the low
dimensionality of the ordered phase. Due to frustration the ordering
may only occur on a low dimensional scale: for instance in three
dimensions each layer has an independent internal ordering. Thus we
expect a larger $\beta_c$ for $d>2$ since the increasing of $d$ only
introduces larger fluctuations, with each molecule having $(d-2)$
allowed out-of-plane orientations. For $d=3$ the low temperature
phase can be regarded as a quenched disordered superposition of
layers which are internally ordered along different in-plane
directions. Thus as a consequence of frustration the system shows
a two-dimensional character below the critical point while behaving as
truly three-dimensional in the high temperature domain.

We should wonder if the MF approach is reliable and how much this
description is affected by the MF approximation. In fact MF associates
the existence of long range order with some broken-symmetry anisotropic
field. On the other hand frustration enforces the long range order to
be anisotropic in space: this is exact since we have shown that, at
variance with usual magnetic models, even the ground state does not
have an isotropic long range order. Long range order may only occur
inside low-dimensional sub-sets of the lattice. Such sub-sets may 
even be one-dimensional, as infinite polymeric chains of aligned
molecules. For $d=2$ the existence of such chains is equivalent to
the existence of two-dimensional long range order; for $d>2$ the
existence of infinitely long one-dimensional chains does not imply the
existence of a two-dimensional order, since the chains could be
randomly oriented along all the lattice axes. In such sense long
range order does not necessarily imply the breaking of isotropy.
Any configuration characterized by one-dimensional long range order
would have an higher energy compared to the ground state, and
could be stable only in an intermediate range of temperature. We
expect that decreasing the temperature the generic d-dimensional
system could reach first a one-dimensional order and then the full 
two-dimensional
order (i.e. order inside each two-dimensional layer). This already
corresponds to a minimum of the configurational energy, since 
nothing is
gained by extending the ordering to higher dimensional sub-sets.
The possible existence of an intermediate isotropic 
phase characterized by one-dimensional long range order is an open
question which cannot be addressed by the present MF approach. The
critical temperature should be higher as compared to full ordering
since a free one-dimensional chain of molecules, each having $d$
allowed orientational states, always shows long range order. By an
argument similar to that employed below eq. (4), such one-dimensional
chain can be mapped on a one-dimensional magnetic model in presence
of a non-zero magnetic field, and is characterized by a non-zero
net magnetization for any value of $\beta$.

Such speculation is corroborated by several estimates of $\beta_c$
obtained for $d=3$ by a modified Migdal-Kadanoff renormalization
group approach [9] or by employing an approximate duality relation [10].
In both cases no anisotropic field is required, and the transition
points are detected through the singular behaviour of the free energy
at the onset of long range order. Details about such techniques will be
published elsewhere. We may anticipate the prediction $\beta_c\approx
0.15$ for $d=3$, to be compared to the MF result $\beta_{MF}=3/8=0.375$.
Since any improving of the MF approach would eventually yield an
increase of $\beta_{MF}$ as a consequence of fluctuations [10], we are
tempted to attribute the different estimates for $\beta_c$ to two
different transitions. Of course further work is required in order
to establish the possible existence of such intermediate phase.

In summary a new non-trivial extension of the Ising model has been
proposed for $d\ge 2$, incorporating frustration effects and thus
reproducing some aspects of ordering in covalent molecular systems.
The model undergoes an order-disorder transition from an high 
temperature
isotropic state to a low temperature anisotropic low-dimensional phase.
A MF analysis is presented for a generic $d\ge 2$, and the results have
been discussed in comparison to some exact limit properties of the
$d$-dimensional model. The possible existence of an intermediate
isotropic phase, characterized by one-dimensional long range order,
is proposed as an open question for further work on the model.
}
\vfill\eject

{\ner
REFERENCES}

\vskip 15pt

{\tit
\item {[1]} I.I.Mazin, R.J.Hemley, A.F.Goncharov, M.Hanfland, H.K.Mao,
preprint Cond-mat/9605083.
\item {[2]} G.Piccitto, private communication.
\item {[3]} F.Siringo, R.Pucci, N.H.March, {\it High Pressure Res.} 
{\bf 2}, 109 (1990).
\item {[4]} R.Pucci and F.Siringo, in {\it Molecular Systems Under High
Pressure}, R.Pucci and G.Piccitto Eds., North-Holland (1991).
\item {[5]} M.Pasternak, J.N.Farrell, R.D.Taylor, {\it Phys.Rev.Lett.}
{\bf 58}, 575 (1987). 
\item {[6]} F.Siringo, R.Pucci, N.H.March, 
{\it Phys.Rev.B} {\bf 37}, 2491 (1988).
\item {[7]} F.Siringo, R.Pucci, N.H.March, {\it Phys.Rev.B} 
{\bf 38}, 9567 (1988).
\item {[8]} For a review see for instance: C.Itzykson,J.-M.Drouffe,
{\it Statistical Field Theory}, Cambridge Univ.Press (1989).
\item {[9]} F.Siringo, preprint (1996).
\item {[10]} F.Siringo, unpublished.
\item{}
}

\vfill\eject

{\ner
Figure Captions}

\vskip 15pt

{\tit
\item{Fig.1} An allowed configuration for $d=2$.

\item{Fig.2} Ground state configurations for the two-dimensional
{\it attractive} (a) and {\it repulsive} (b) models.
\item{}
}

\vfill\eject

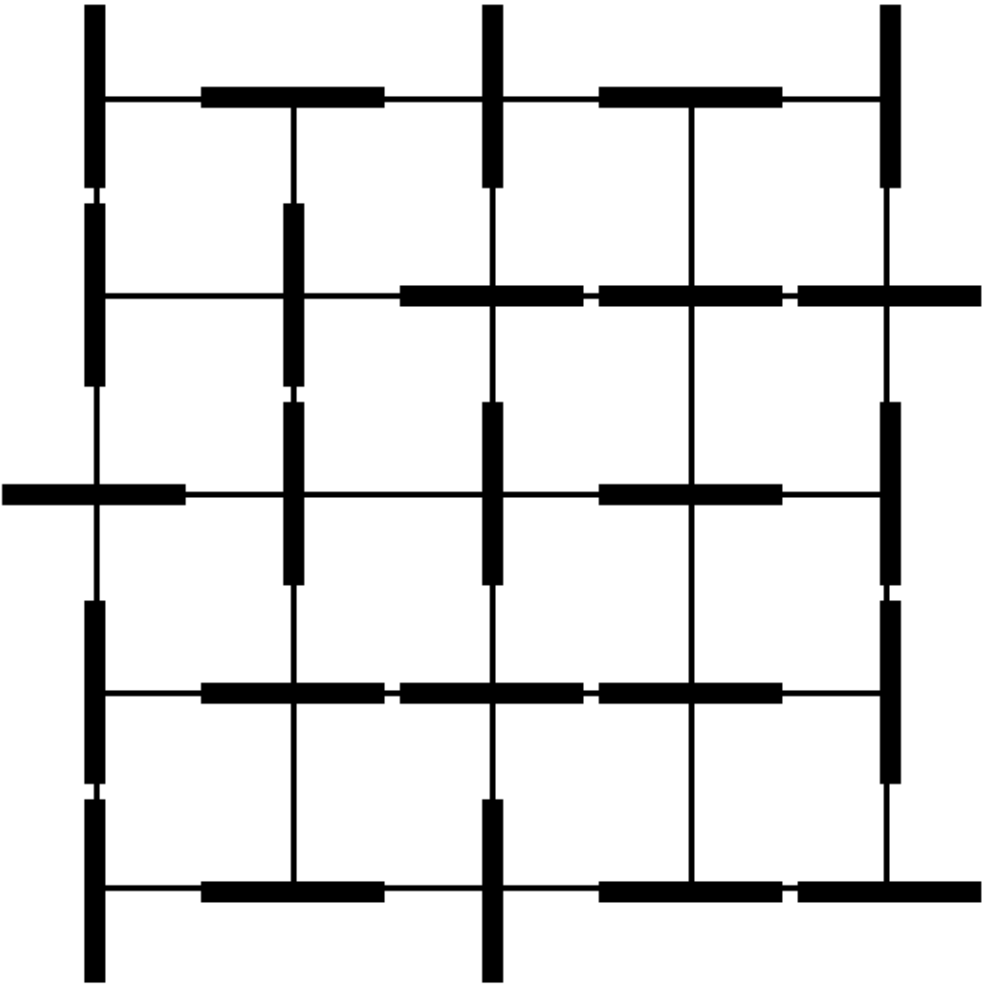
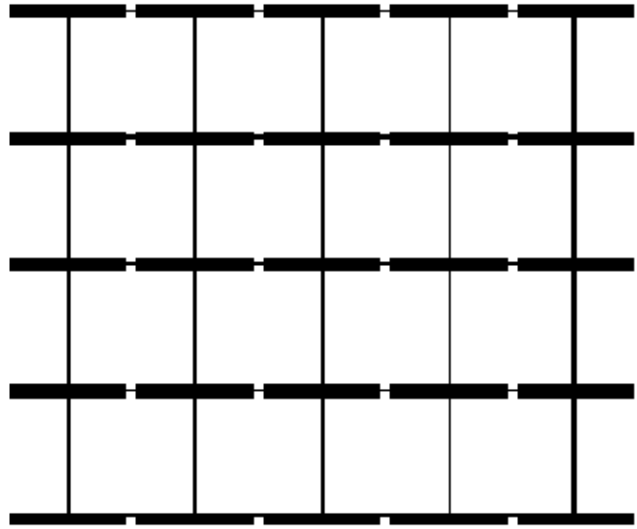
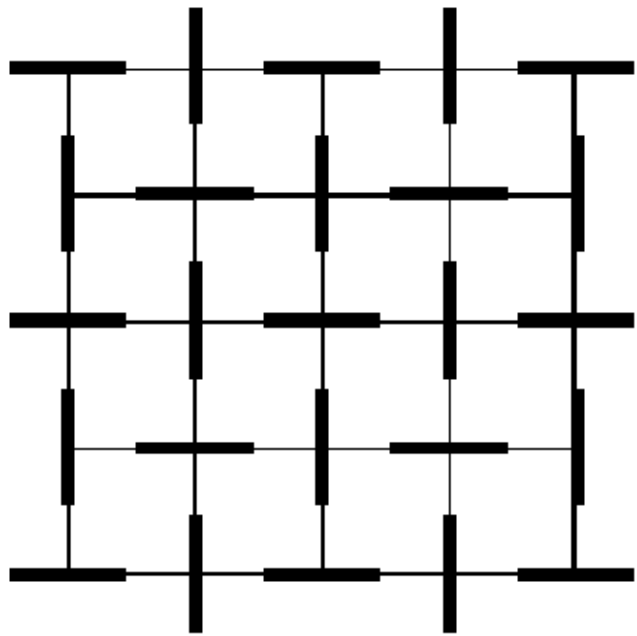
\bye